\documentclass{article}
\usepackage{amssymb,amsmath}

\newcommand{\la}{\lambda}
\newcommand{\La}{\Lambda}
\newcommand{\ve}{\varepsilon}

\newcommand{\de}{\delta}
\newcommand{\De}{\Delta}

\newcommand{\nn}{\nonumber}
\newcommand{\x}{\hat{x}}

\newcommand{\lf}{\ell}

\newcommand{\hR}{\hat{R}}
\newcommand{\trr}{\triangleright}
\newcommand{\ck}{\check}
\newcommand{\cD}{{\cal D}}

\def\ha{\frac{1}{2}}

\def\Ax{{\mathcal{A}_{\x}}}

\def\ds{\stackrel{*}{,}}

\def\pat{\partial}

\begin{document}

\begin{titlepage}
\rightline{LBNL-51361}
\rightline{MPI-PhT/2002-39}

\vspace{4em}
\begin{center}

{\Large{\bf Non-Abelian Gauge Theory \\ on
    \boldmath $q$-Quantum Spaces}}

\vskip 3em

{{\bf
S.\ Schraml}}

\vskip 1em

   Theoretical Physics Group, Bldg. 50A5104\\
   Lawrence Berkeley National Laboratory\\
   One Cyclotron Road, Berkeley, CA 94720\\[1em]
   and\\[1em]
   Max-Planck-Institut f\"ur Physik\\
   F\"ohringer Ring 6, D-80805 M\"unchen\\[1em]

\end{center}

\vspace{2em}

\begin{abstract}
Gauge theories on $q$-deformed spaces are constructed 
using covariant derivatives. For this purpose a ``vielbein''
is introduced, which transforms under gauge transformations. 
The non-Abelian case is treated by establishing a connection 
to gauge theories on commutative spaces, i.e. by a 
Seiberg-Witten map. As an example we consider the Manin 
plane. Remarks are made concerning the relation between 
covariant coordinates and covariant derivatives.
\end{abstract}

\vfill
\noindent \hrule\vskip.1cm
\hbox{{\small{\it e-mail: }}{\small\quad
SLSchraml@lbl.gov}}
\end{titlepage}\vskip.2cm

\newpage
\setcounter{page}{1}

\section{Introduction}

Gauge theories on noncommutative spaces have recently found some 
interest in mathematical high energy physics. Most of the work 
concentrated on noncommutative spaces, for which the commutator 
of two coordinates is constant: 
$[\x^i,\x^j]=i\theta^{ij}\in\mathbb{C}$. 
From the point of view of gauge theory this space is nice, since 
the algebra relations of the derivatives do not get deformed 
(up to possible constants in the commutator of two derivatives). 
This allows for example to write for the gauge 
transformation of the components of the gauge potential 
$\delta A_i=\pat_i\La+i\La*A_i-iA_i*\La$; i.e. only the noncommutative 
$*$-product has to be used instead of the commutative point-wise 
product. The reason for this is the unchanged Leibniz rule. 

However, for more general noncommutative spaces the derivatives 
-- if they exist -- and also the Leibniz rule change. 
A case for which differential calculi 
have been worked out is the structure of $q$-deformed quantum spaces 
\cite{woron-diffcalc,cov_diff}. 
It has the advantage, that not only the space is deformed, but also 
the symmetries of the space get deformed. 
In this paper we treat gauge theory on these spaces by introducing 
covariant derivatives. 
The case of a  general non-Abelian gauge group is dealt with 
by constructing a Seiberg-Witten map, i.e. a map which connects 
the gauge theory on the noncommutative space with gauge theory 
on a commutative space \cite{SWi}. For this purpose we use the 
$*$-product 
formalism; in particular the derivatives become in this formulation 
power series in the deformation parameter.

We start with recalling some definitions and properties of 
$q$-deformed spaces \cite{drinfeld,woron-qgroups,FRT}, 
mainly to fix the notation (which we took 
form \cite{Schmdg}). 
We think of quantum spaces as being defined by the 
commutation relations of their 
coordinates $\x^1,\ldots,\x^N$, which are quadratic in our case:
\begin{equation}
  \label{xx-rel}
  P_-{}^{ij}_{kl}\x^k\x^l=0,\qquad P_-{}^{ij}_{kl}\in\mathbb{C}.
\end{equation}
This means, that we consider the algebra of formal power series in 
the elements $\x^i$ modulo the ideal generated by the relations 
(\ref{xx-rel}):
\begin{equation}
  \Ax\equiv \frac{\mathbb{C}[[\x^1,\ldots,\x^N]]}{P_-\x\x}
\end{equation}
as defining the ``noncommutative space''. The matrix $P_-$ is the 
($q$-deformed) anti-symmetric projector appearing in the 
projector decomposition 
of the $\hR$-matrix of the respective quantum group:
\begin{eqnarray}
  GL_q, SL_q &:&\hR=qP_+-q^{-1}P_-\nn\\
  SO_q, SP_q &:&\hR=qP_+-q^{-1}P_-+q^{1-N}P_0
\end{eqnarray}
with $\sum P=1$; the $\hR$-matrix fulfills the 
Yang-Baxter equation:
\begin{equation}
    \hR_{12}\,\hR_{23}\,\hR_{12}\,=\,\hR_{23}\,\hR_{12}\,\hR_{23},
\end{equation}
where we used the notation $\hR^{\quad i_1 \, i_2 \, i_3}_{12 \, 
j_1 \, j_2 \, j_3} \equiv \hR^{i_1 \, i_2}_{j_1 \, j_2}\, 
\delta^{i_3}_{j_3}$ and $\hR^{\quad i_1 \, i_2 \, i_3}_{23 \, 
j_1 \, j_2 \, j_3} \equiv \delta^{i_1}_{j_1}\, \hR^{i_2 \, 
i_3}_{j_2 \, j_3}$. 
The parameter $q=e^h\in \mathbb{C}$ is the deformation 
parameter; for $q\to 1$ one recovers the undeformed 
commutative space. 
Given the $\hR$-matrix, the quantum group $G_q$ itself is 
defined by the relations of 
its (noncommutative) algebra of functions, which is generated 
by $u_i^j;\, i,j:1,\ldots, N$:
\begin{equation}
  \hR^{ij}_{kl}u^k_mu^l_n=u^i_ku^j_l\hR^{kl}_{mn}.
\end{equation}
The noncommutative algebra $\Ax$ is a comodule algebra 
of the respective quantum group; and it is a module algebra of 
the dual of the quantum group, namely of the deformed enveloping 
algebra $U_q(g)$ of the Lie-algebra related to the respective 
group. The deformed enveloping algebra is generated by the so 
called $L$-functionals $\hat\lf^+_{ij}$ and $\hat\lf^-_{ij}$, 
$i,j=1\ldots N$, which satisfy 
\begin{equation}
  \hR^{lk}_{ij}\hat\lf^{\pm}_{ks}\hat\lf^{\pm}_{lt}=
  \hat\lf^{\pm}_{jl}\hat\lf^{\pm}_{ik}\hR^{ts}_{kl}.
\end{equation}
and for which holds: $\hat\lf^+_{ij}=0$ for $j<i$, $\hat\lf^-_{ij}=0$ 
for $j>i$.

There exist two well known covariant, w.r.t. the (co-)action of 
the quantum group, differential calculi for these 
quantum spaces:
\begin{eqnarray}
  \label{x-dx-rel}
  d\x^id\x^j&=&-q^{\pm 1}\hR^{\pm 1}{}^{ij}_{kl}d\x^kd\x^l\nn\\
  \x^id\x^j&=&q^{\pm 1}\hR^{\pm 1}{}^{ij}_{kl}d\x^k\x^l  
\end{eqnarray}
Here the exponents $+1$ and $-1$ correspond to the first and 
the second differential calculus respectively. 
The associated derivatives $\hat\pat_i$ ($d=d\x^i\hat\pat_i$) fulfill
\begin{eqnarray}
  \label{dd-rel}
  P_-{}^{ij}_{kl}\hat\pat_j\hat\pat_i&=&0\\  
  \hat\pat_i\x^j&=&\de_i^j+q^{\pm 1}
  \hR^{\pm 1}{}^{jl}_{ik}\x^k\hat\pat_l\nn
\end{eqnarray}
Since in the $q\to1$ limit $\hR^{jl}_{ik}$ goes to 
$\delta^j_k\delta^l_i$, this is clearly a deformation 
of the usual differential calculus. 
Defining
\begin{equation}
  P_S\equiv 1-P_-
\end{equation}
the commutation relations among the differentials can be written as
\begin{equation}
  {P_S}^{kl}_{ij}d\x^id\x^j=0.
\end{equation}

\section{Realization on functions}\label{funcReal}

In the following we realize the whole noncommutative algebra, 
including coordinates $\x$, partial derivatives $\hat\pat$ and the 
deformed enveloping algebra $U_q(g)$, as operators acting on the 
noncommutative coordinate algebra $\Ax$. I.e. we ask for operators on
the algebra $\Ax$ of coordinates, such that the above relations hold.
To distinguish between the elements $\hat a$ of the algebra and the
corresponding
mappings, we will denote the latter by $\ck a$:
\begin{equation}
  \ck a:\Ax\longrightarrow\Ax.
\end{equation}

For explicit formulas we use a basis of the algebra $\Ax$. 
Due to the Yang-Baxter equation such a basis is for example 
given by ordered monomials in the case of $q$-deformed 
quantum spaces :
\begin{equation}
  \label{xbasis}
  \{(\x^1)^{i_1}(\x^2)^{i_2}\cdots (\x^N)^{i_N}|i_n=0,1,2,\ldots\}.
\end{equation}

Clearly, the coordinates $\x^i$ act on functions by multiplication:
\begin{equation}
   (\ck x^i f)(\x)\equiv \x^if(\x).
\end{equation}
In terms of a basis this means essentially a reordering.

The quantum group $G_q$ coacts on coordinates, which on the 
other hand also form a vector
representation of the deformed enveloping algebra $U_q(g)$. Using
the $L$-functionals $\hat\lf^{\pm}_{ij}$ as generators of $U_q(g)$ these 
operations are connected:
\begin{equation}
  \label{Lx-action}
  \hat\lf^{\pm}_{ij}\trr \x^k=\langle\hat\lf^{\pm}_{ij},u_k^l\rangle
\x^l=\hat{R}^{\pm 1}{}^{kj}_{il}\x^l.
\end{equation}
Here we have used the right coaction 
$\x^m\mapsto\x^l\otimes u_m^l$,
and the dual pairing $\langle\cdot,\cdot\rangle: 
G_q\times U_q(g) \to\mathbb{C}$, 
which gives rise to a left action $\trr$, and which is 
related to the action according to the vector representation.
``Related'' means that there might be an additional $q$-factor, if the
$R$-matrix is not exactly given by the $R$-matrix $R_{VV}$ of the
fundamental vector representation of the $q$-deformed enveloping
algebra.
This is the case in the example we treat below, in which
$R=q^{\ha}R_{VV}$.
Since the coproduct of $U_q(g)$ is known:
\begin{equation}
  \label{delta-lf}
  \Delta(\hat\lf^{\pm}_{ij})
  =\hat\lf^{\pm}_{ik}\otimes\hat\lf^{\pm}_{kj},
\end{equation}
the action of the operators $\ck\lf^{\pm}_{ij}$, i.e. 
of $U_q(g)$, on functions (polynomials in the coordinates $\x^i$) 
is determined.

The commutation relations between the coordinates and the
$L$-functionals are obtained by a crossed product 
construction, that is $a\x=\sum (a_{(1)}\trr\x)a_{(2)}$ 
in our case: 
\begin{equation}
  \hat\lf^{\pm}_{ik}\x^m
  =\hat{R}^{\pm 1}{}^{mj}_{il}\x^l\hat\lf^{\pm}_{jk}.
\end{equation}
Here $\hR^{+1}$ appears in the $\hat\lf^+\x$-relations and 
$\hR^{-1}$ in the $\hat\lf^-\x$-relations, 
cf. Eqn. (\ref{Lx-action}). 
Again, there might be an additional factor in front of $\hR$
according to the relation between $R$ and $R_{VV}$. 
With this at hand the action on a polynomial in $\x$ can be
obtained as follows: multiply the function from the left with some
$\hat\lf^{\pm}_{ij}$, commute this $\hat\lf^{\pm}_{ij}$ and all 
$\hat\lf^{\pm}$
obtained during this process to the whole right,
set all $\hat\lf^{\pm}_{kk}$ equal to $1$ and skip all terms with a
$\hat\lf^{\pm}_{kl},k\neq l$ on the right.

For the partial derivatives we first determine the commutation
relations between differentials $d\x^i$ and functions $f(\x)$.
We treat the first differential calculus 
and start with relation (\ref{x-dx-rel}):
\begin{eqnarray}
  \x^id\x^j&=&q\hat{R}^{ij}_{kl}d\x^k\x^l\nn\\
  &=&q\langle \hat\lf^+_{kj},u_i^l\rangle d\x^k\x^l\nn\\
  &=&qd\x^k\hat\lf^+_{kj}\trr \x^i.
\end{eqnarray}
To keep track of the $q$ on the right hand side we extend our algebra
by a scaling operator\footnote{It corresponds to the
scaling operator which one uses
to construct real momentum operators \cite{ZO,LWW}.}
$\hat U$ for which holds:
\begin{equation}
  \hat U\x^i=q\x^i\hat U,\quad \hat U\hat\pat_i=q^{-1}
  \hat\pat_i\hat U,\quad\hat U\hat\lf^{\pm}_{ij}
  =\hat\lf^{\pm}_{ij}\hat U.
\end{equation}
On functions the scaling operator is realized by
\begin{eqnarray}
  && \ck U f(\x)=f(q\x)\nn\\
  && \ck U(fg)=(\ck Uf)(\ck Ug).
\end{eqnarray}
If we take into account the coproduct (\ref{delta-lf}) of the
$L$-functionals, or in other words the quasitriangularity of the 
deformed enveloping algebra, we arrive at
\begin{equation}
  f(\x)d\x^j=d\x^k(\hat U\hat\lf^+_{kj})\trr f(\x)
  \equiv d\x^k (B_k{}^jf(\x)).
\end{equation}
Where we defined the operator 
\begin{equation}
  \label{B-def}
  B_i{}^j=\ck U\ck\lf^+_{ij},
\end{equation}
which fulfills
\begin{equation}
  \label{Bcoprod}
   B_i{}^j(fg)=( B_i{}^kf)( B_k{}^jg)
\end{equation}
and
\begin{equation}
  \label{BB-rel}
  \hat{R}^{lk}_{ij}B_k{}^sB_l{}^t=B_j{}^lB_i{}^k
  \hat{R}^{ts}_{kl}
\end{equation}
due to the same relations for the $L$-functionals. 
This means, that the $B_i{}^j$ form a realization of 
the quantum group. 

If we had started with the second differential calculus, 
we would have obtained 
$B_i{}^j=\ck U^{-1}\ck\lf^-_{ij}$, where we extended 
the algebra by the inverse of $U$.

Applying $d=d\x^i\hat\pat_i$ to a function we define 
the realization $\ck\pat_i$ 
of the partial derivatives on functions $f$ as follows:
\begin{equation}
  \label{deriv-commute}
  d f=d\x^i[(\ck\pat_if)+(B_i{}^jf)\hat\pat_j].
\end{equation}
We obtain therefore the Leibniz rule
\begin{equation}
  \label{leibniz}
  \ck\pat_i(fg)=(\ck\pat_if)g+(B_i{}^jf)(\ck\pat_jg).
\end{equation}
Eqn. (\ref{Bcoprod}) ensures that 
$\ck\pat_i(f(gh))=\ck\pat_i((fg)h)$,
which must hold since the algebra $\Ax$ of coordinates 
is associative.

Eqn. (\ref{deriv-commute}) shows, that the explicit 
formulas for the action of the derivatives $\ck\pat_i$ 
on functions $f(\x)$ can be obtained in the same way
as described above for the $L$-functionals. Namely, 
commute $\hat\pat_i$ from the left side to the right 
side of $f(\x)$ and forget about terms
with some derivative $\hat\pat$ on the right.

Since the differential calculus is covariant
w.r.t. the quantum group, 
we obtain for the commutation relation
between the $L$-functionals and the derivatives:
\begin{equation}
  \hat\pat_m\hat\lf^{\pm}_{lj}=\hat{R}^{\pm 1}{}^{ik}_{lm}
  \hat\lf^{\pm}_{kj}\hat\pat_i.
\end{equation}
As above for the coordinates, we have 
$\hat\pat\hat\lf^+=\hR^{+1}\hat\lf^+\hat\pat$ 
and $\hat\pat\hat\lf^-=\hR^{-1}\hat\lf^-\hat\pat$ 
respectively. 
Using Eqn. (\ref{B-def}) we therefore obtain: 
\begin{equation}
  \ck\pat_mB_l{}^j=q\hat{R}^{ik}_{lm}B_k{}^j\ck\pat_i.
\end{equation}

The action on products of functions (\ref{leibniz}) can be 
interpreted as comultiplication:
\begin{eqnarray}
  \Delta(\pat_i)&=&\pat_i\otimes 1+B_i{}^j\otimes\pat_j\\
  \Delta(U)&=&U\otimes U\nn
\end{eqnarray}
One can show, that this together with (\ref{delta-lf}) and 
the counit
\begin{equation}
  \ve(\lf^{\pm}_{ij})=\de_{ij},\quad\ve(\pat_i)=0,
  \quad\ve(U^{\pm 1})=1
\end{equation}
defines a bialgebra structure on the algebra generated by
$\lf^{\pm}_{ij}$, $\pat_i$ and $U^{\pm 1}$.
Since the action of the antipode of $U_q(g)$ on the 
$L$-functionals is known, we can equip this bialgebra 
easily with a antipode $S$ to get a Hopf algebra:
\begin{equation}
  S(U^{\pm1})=U^{\mp 1},\quad S(\pat_i)=-S(B_i{}^k)\pat_k.
\end{equation}
Indeed, in this way we get two bialgebra structures, 
since we have either $B_i{}^j=U\lf^+_{ij}$ or 
$B_i{}^j=U^{-1}\lf^-_{ij}$, depending on 
which differential calculus we use. The relations among
$\lf^{\pm}_{ij}$, $\pat_i$ and $U^{\pm 1}$ are the same 
in both cases. However, the algebra
$\Ax$ of  coordinates, on which this algebra acts in 
two different ways, is only a module algebra of this 
bialgebra if one chooses the appropriate comultiplication. 

Below we will connect the gauge theory on the 
noncommutative space with gauge theory on commutative 
spaces. This will be done using 
$*$-products \cite{BFFLS}.
For the construction of a $*$-product resembling a given 
algebra $\Ax$ we use the basis for which we have obtained  
the action of all operators above. 
To be specific we take the basis (\ref{xbasis}), i.e. 
normal ordering. We define a map 
$W$ (``quantization'') from
the algebra of $N$ commuting variables 
$x^1,\ldots,x^N$ to $\Ax$ by
\begin{equation}
  W\big((x^1)^{i_1}\cdots(x^N)^{i_N}\big)\equiv
  (\x^1)^{i_1}\cdots(\x^N)^{i_N}.
\end{equation}
The requirement, that this map is an algebra 
homomorphism defines a 
$*$-product for polynomials of commuting variables:
\begin{equation}
  W(f*g)\equiv W(f)W(g).
\end{equation}
Since in the case of quantum spaces $W$ is clearly 
invertible
(due to the Poincar\'e-Birkhoff-Witt property), 
the $*$-product exists. 
This $*$-product can be written in terms of a 
formal power series of 
differential operators with formal 
parameter $h\equiv\ln q$. 
Formally we can extend it to 
$C^{\infty}(\mathbb{R}^N)[[h]]$.

Furthermore we are able to translate not only the 
product of functions, 
but also operators acting on the noncommutative 
algebra $\Ax$. Given some operator $\ck O$
on $\Ax$ this is done by
\begin{equation}
  W(\ck Of)\equiv \ck OW(f).
\end{equation}
Therefore we get for all the operators 
$\ck\lf^{\pm}_{ij},\ck\pat_i,\ck U^{\pm 1}$
defined above, corresponding differential operators 
acting on $C^{\infty}$-functions on $\mathbb{R}^N$. 
These operators are power series in the deformation 
parameter (and fulfill by definition the 
algebra relations of $\lf^{\pm}_{ij},\pat_i,U^{\pm 1}$).

As an illustration we consider 
the $q$-deformed two-dimensional plane (Manin-plane), 
which has a $SL_q(2)$-symmetry. 
For more complicated examples see \cite{Claudia}. 
The well known $\hR$-matrix is given by
\begin{equation}
  \label{R-matrix}
  \hR=\left(
    \begin{array}[]{cccc}
      q&0&0&0\\
      0&\la&1&0\\
      0&1&0&0\\
      0&0&0&q
    \end{array}
  \right)
\end{equation}
where $\la=q-q^{-1}$. If we consider the first 
differential calculus ($\x^id\x^j=q\hR_{kl}^{ij}d\x^k\x^l$) 
we obtain the following relations:
\begin{eqnarray}
  &&\x^1\x^2=q\x^2\x^1\\
  &&\hat\pat_1\hat\pat_2=q^{-1}\hat\pat_2\hat\pat_1\nn\\
  &&\hat\pat_1\x^1=1+q^2\x^1\hat\pat_1+q\la\x^2
\hat\pat_2,\qquad\hat\pat_1\x^2=q\x^2\hat\pat_1\nn\\
  &&\hat\pat_2\x^1=q\x^1\hat\pat_2,\qquad\hat\pat_2\x^2=1+q^2\x^2
\hat\pat_2\nn
\end{eqnarray}
For the relations between the coordinates and the 
$U_q(sl_2)$-generators 
one has to take into account
a $q^{\ha}$ due to the relation between the $R$-matrix 
(\ref{R-matrix}) and the $R$-matrix 
for the vector representation $R_{VV}$. 
This implies $\hat U\x^i=q^{\frac{3}{2}}\x^i\hat U$; using the 
second differential calculus would however require a scaling by 
$q^{-\frac{1}{2}}$. 
We obtain ($\lf^+_{21}=0$):
\begin{eqnarray}
  &&\hat\lf^+_{11}\x^1=q^{\ha}\x^1\hat\lf^+_{11},
  \quad\hat\lf^+_{12}\x^1=q^{-\ha}(q\x^1\hat\lf^+_{12}
  +\la\x^2\hat\lf^+_{22}),\quad\hat\lf^+_{22}\x^1
  =q^{-\ha}\x^1\hat\lf^+_{22}\nn\\
  &&\hat\lf^+_{11}\x^2=q^{-\ha}\x^2\hat\lf^+_{11},
  \quad\hat\lf^+_{12}\x^2=q^{-\ha}\x^2\hat\lf^+_{12},
  \quad\hat\lf^+_{22}\x^2=q^{\ha}\x^2\hat\lf^+_{22}
\end{eqnarray}

For the explicit realization on functions we take the basis
$(\x^1)^n(\x^2)^m$. The, in this case simple, calculations
yield:
\begin{eqnarray}
  \label{man-abl-wirk}
  \ck\pat_1\left((\x^1)^n(\x^2)^m\right)&=&
  q^{n+2m-1}[n](\x^1)^{n-1}(\x^2)^m\\
  \ck\pat_2\left((\x^1)^n(\x^2)^m\right)&=&
  q^{n+m-1}[m](\x^1)^{n}(\x^2)^{m-1}\nn\\
  \ck\lf^+_{11}\left((\x^1)^n(\x^2)^m\right)&=&
  q^{\ha(n-m)}(\x^1)^n(\x^2)^m\nn\\
  \ck\lf^+_{12}\left((\x^1)^n(\x^2)^m\right)&=&
  \la q^{\ha(m-n)}[n](\x^1)^{(n-1)}(\x^2)^{(m+1)}\nn\\
  \ck\lf^+_{22}\left((\x^1)^n(\x^2)^m\right)&=&
  q^{-\ha(n-m)}(\x^1)^n(\x^2)^m\nn\\
  \ck\lf^{-}_{21}\left((\x^1)^n(\x^2)^m\right)&=&
  -\la q^{\frac{1}{2}(n-m)+1}[m](\x^1)^{(n+1)}(\x^2)^{(m-1)}\nn\\
  \ck U\left((\x^1)^n(\x^2)^m\right)&=&q^{\frac{3}{2}(n+m)}(\x^1)
^n(\x^2)^m\nn
\end{eqnarray}
In these equations $\la=q-q^{-1}$ and 
$[n]=\frac{q^n-q^{-n}}{q-q^{-1}}$ are defined as usual.

Using the basis $(\x^1)^n(\x^2)^m$, $n,m\in\mathbb{N}$, 
the $*$-product is found to be \cite{MSSW}
\begin{equation}
    f*g = \left. q^{-y^1\frac{\pat}{\pat y^1}x^2
        \frac{\pat}{\pat x^2}}f(x^1,x^2)g(y^1,y^2)
    \right|_{\genfrac{}{}{0pt}{}{y^1\to x^1}{y^2\to x^2}}
\end{equation}
In the case of the higher-dimensional $GL_q$- and 
$SL_q$-quantum 
planes the $*$-product is quite simple and very similar to 
the 2-dimensional. For Euclidian quantum spaces however the 
formulas become complicated \cite{WW}. 

The derivatives $\ck\pat_i$ and the $B_i{}^j$ operators become 
($\pat_i=\frac{\pat}{\pat x^i}$):
\begin{eqnarray}
  \label{Man-*Wirk}
  \ck\pat_1f&=&\frac{1}{\la x^1}
  q^{2x^2\pat_2-1}(q^{2x^1\pat_1}-1)\,f\nn\\
  \ck\pat_2f&=&\frac{1}{\la x^2}
  q^{x^1\pat_1-1}(q^{2x^2\pat_2}-1)\,f\nn\\
  \ck B_1{}^1f&=&q^{2x^1\pat_1+x^2\pat_2}\,f\nn\\
  \ck B_1{}^2f&=&\frac{x^2}{x^1}q^{2x^2\pat_2}
  (q^{2x^1\pat_1}-1)\,f\nn\\
  \ck B_2{}^2f&=&q^{x^1\pat_1+2x^2\pat_2}\,f
\end{eqnarray}
Using $\frac{1}{x}f(x\pat_x)=f(x\pat_x+1)\frac{1}{x}$, 
it is easy to check explicitly, that the algebra 
relations are really satisfied. I.e. the algebra is 
realized by differential operators acting on 
$C^{\infty}$-functions on $\mathbb{R}^N$. 

\section{Covariant derivatives}

In the following we treat gauge theory on quantum spaces by 
constructing covariant derivatives. 

We assume that an infinitesimal gauge transformations of a
field $\psi(\x)$, which belongs to a representation of 
the gauge group, is given by:
\begin{equation}
  \de\psi(\x)=i\La(\x)\psi(\x).
\end{equation}
One might think of the gauge parameter $\La(\x)$ as a 
matrix and of $\psi(\x)$ as a column vector, 
both with entries in $\Ax$; see below, 
section \ref{sect_swmap}.

We require $\de(\cD_i\psi)=i\La(\cD_i\psi)$ 
for the covariant derivation of a field and 
make for this purpose the ansatz:
\begin{equation}
  \label{covDer}
  {\cD}_i\psi\equiv E_i{}^j(\ck\pat_j-iA_j)\psi.
\end{equation}
Where we introduced a gauge field $A_i$ and a ``vielbein'' 
field $E_i{}^j$ which we assume to be invertible:
\begin{equation}
  E_i{}^ke_k{}^j=\de_i^j=e_i{}^kE_k{}^j.
\end{equation}
From the requirement of covariance under gauge 
transformations we obtain:
\begin{eqnarray}
\label{gauge-trafo}
  &&\de E_i{}^j=i\La E_i{}^j-iE_i{}^k(B_k{}^j\La)\nn\\
  &&\de A_i=\ck\pat_i\La+i(B_i{}^j\La)A_j-iA_i\La
\end{eqnarray}
The inverse of the vielbein transforms according to
\begin{equation}
  \de e_i{}^j=i(B_i{}^k\La)e_k{}^j-ie_i{}^j\La.
\end{equation}

For $j<i$ the operators $B_i{}^j$ vanish for the above choice 
(\ref{B-def}). Therefore it is compatible with the  
transformation law (\ref{gauge-trafo}) to choose 
$E_i{}^j\equiv 0\equiv e_i{}^j$ for $j<i$. 
With this choice it is easy to express $e_i{}^j$ in terms of 
$E_i{}^j$. 

To obtain tensors we apply two covariant derivatives to 
a field. Using the definition (\ref{covDer}) and the
relation (\ref{dd-rel}) for the partial derivatives, 
we arrive at
\begin{equation}
  \cD_i\cD_j\psi=G_{ij}{}^{kl}\cD_k\cD_l\psi+T_{ij}{}^k
  \cD_k\psi+F_{ij}\psi.
\end{equation}
Where $G_{ij}{}^{kl}$, the torsion $T_{ij}{}^k$ and the field
strength $F_{ij}$ transform as tensors:
\begin{equation}
  \de G_{ij}{}^{kl}=i[\La,G_{ij}{}^{kl}],\quad\de T_{ij}{}^k
  =i[\La,T_{ij}{}^k],\quad\de F_{ij}=i[\La,F_{ij}].
\end{equation}
Explicitly one finds the following expressions for 
these tensors:
\begin{eqnarray}
  \label{tensors}
  G_{ij}{}^{kl}&=&E_i{}^m(B_m{}^rE_j{}^n)
  P_S{}_{nr}{}^{ts}(B_s{}^ue_t{}^l)e_u{}^k\nn\\
  T_{ij}{}^k&=&\left[(\cD_iE_j{}^m)-G_{ij}{}^{rn}
  (\cD_rE_n{}^m)\right]e_m{}^k\nn\\
  iF_{ij}&=&E_i{}^k(B_k{}^rE_j{}^n)
  \big[(\ck\pat_rA_n)+i(B_r{}^uA_n)A_u\nn\\
  &&\qquad\qquad\qquad-P_S{}_{nr}{}^{ts}
  \left((\ck\pat_sA_t)+i(B_s{}^uA_t)A_u\right)\big]\nn\\
  &=&E_i{}^k(B_k{}^rE_j{}^n)P_-{}_{nr}^{ts}
  (\ck\pat_sA_t+i(B_s{}^uA_t)A_u)
\end{eqnarray}
The covariant derivative $\cD_iE_j{}^k$ of the vielbein, 
which appears in the expression for the torsion 
$T_{ij}{}^k$, turns out to be
\begin{equation}
  \cD_iE_j{}^k=E_i{}^m\left[(\ck\pat_m-iA_m)E_j{}^k
  +iq(B_m{}^rE_j{}^n)\hR_{nr}{}^{ts}(B_s{}^kA_t)\right].
\end{equation}
The zeroth order contribution in the expansion of $F_{ij}$ in 
powers of $h$ clearly reduces to its usual commutative 
counterpart, at least up to a normalization.

We assume that the field strength is anti-symmetric, where 
anti-symmetric has to be understood in a $q$-deformed sense:
\begin{equation}
  P_-{}_{ji}^{rn}F_{nr}=F_{ij}.
\end{equation}
Then, inspecting the explicit expression for the field 
strength (\ref{tensors}), one finds that 
this requirement is fulfilled, if the following
relation for the vielbein field holds:
\begin{equation}
  \label{E-Bed}
  E_i{}^k(B_k{}^rE_j{}^n)P_-{}_{nr}^{ts}
  =P_-{}_{ji}^{rn}E_n{}^k(B_k{}^sE_r{}^t).
\end{equation}
This condition is covariant under gauge transformations.
Since $P_S=1-P_-$, it holds in the same way for $P_S$.
If it holds the expressions for the
tensors $G_{ij}{}^{kl}$ and $T_{ij}{}^k$ simplify as follows:
\begin{eqnarray}
    G_{ij}{}^{kl}&=&(1-P_-)_{ji}^{lk}\nn\\
    T_{ij}{}^k&=&P_-{}_{ji}^{nr}(\cD_rE_n{}^m)e_m{}^k,
\end{eqnarray}
which leads to
\begin{equation}
  P_-{}_{kl}^{ij}\cD_j\cD_i=T_{lk}{}^m\cD_m+F_{lk}.
\end{equation}

If we define
\begin{equation}
  e^i\equiv d\x^je_j{}^i
\end{equation}
the condition (\ref{E-Bed}) implies on the level of differentials
\begin{equation}
  P_S{}_{ij}^{kl}e^ie^j=0.
\end{equation}

It is also possible to consider finite gauge transformations:
\begin{equation}
  \psi(\x)\mapsto\psi'(\x)=G(\x)\psi(\x).
\end{equation}
The related transformations of the gauge and vielbein fields are 
\begin{eqnarray}
  E'_i{}^j&=&GE_i{}^k(B_k{}^jG^{-1})\nn\\
  A'_i&=&-i(\ck\pat_iG)G^{-1}+(B_i{}^jG)A_jG^{-1}.
\end{eqnarray}
Clearly, this reduces to Eqn. (\ref{gauge-trafo}) for 
$G=1+i\La$ with $\La$ infinitesimal small. 
Accordingly pure gauge fields are given by
\begin{eqnarray}
  E_i{}^j&=&G(B_i{}^jG^{-1})\nn\\
  A_i&=&-i(\ck\pat_iG)G^{-1}=i(B_i{}^jG)(\ck\pat_jG^{-1}).
\end{eqnarray}
For these fields the field strength $F_{ij}$ and the torsion 
$T_{ij}{}^k$ vanish. Also the relation (\ref{E-Bed}) 
is satisfied.

\section{Seiberg-Witten map}\label{sect_swmap}

In this section we connect, for an arbitrary 
non-Abelian gauge group, gauge theories on 
noncommutative spaces with gauge theories on 
commutative spaces \cite{SWi}. This means, 
that we will express the transformation parameter 
$\La$, the gauge field $A_i$ and the vielbein 
$E_i{}^j$ in terms of the commutative 
gauge parameter and field $a_i$. We follow the 
cohomological approach of \cite{zumino,BCZ}. 
For this purpose we introduce a ghost field which  
is a Lie algebra-valued Grassmannian field: 
$c=c_aT^a, \{c_a,c_b\}=0, [T^a,T^b]=if^{ab}{}_cT^c$. 
As usual the BRST transformations of 
the respective fields read:
\begin{eqnarray}
  sc&=&ic^2\nn\\
  sa_i&=&\pat_ic+i[c,a_i]\nn\\
  s\psi&=&ic\psi.
\end{eqnarray}
Where $\pat_i$ is the usual derivative of commuting 
variables. The BRST operator $s$ is nilpotent 
$s^2=0$ and commutes with the $*$-product, the quantum 
derivatives $\ck\pat_i$ and the operators $B_i{}^j$. 

The noncommutative gauge parameter $\La$ is replaced 
by a noncommutative ghost field $C$, which becomes a 
functional of the commutative gauge field $a_i$ and 
ghost field $c$. The relation to be fulfilled by $C$ is 
\begin{equation}
  \label{sw-ghost}
  sC=iC*C
\end{equation}
Once this is solved for $C$, one uses the Eqns:
\begin{eqnarray}
  \label{sw-map-eqns}
  sE_i{}^j&=&iC*E_i{}^j-iE_i{}^k*(B_k{}^jC)\nn\\ 
  sA_i&=&\ck\pat_iC+i(B_i{}^jC)*A_j-iA_i*C\nn\\
  s\psi&=&iC*\psi
\end{eqnarray}
to determine $E_i{}^j$, $A_i$ and $\psi$. 
It is easily seen, that these relations 
are consistent with the nilpotency of the BRST 
operator $s$. The fields $C$, $A_i$, $E_i{}^j$ are 
formal power series in the deformation parameter 
$h\equiv\ln q$: 
\begin{equation}
  C=\sum_{n=0}^{\infty}h^nC^{(n)}[c,a_l],\quad A_i
  =\sum_{n=0}^{\infty}h^nA^{(n)}_i[a_l],\quad E_i{}^j
  =\sum_{n=0}^{\infty}h^nE^{(n)}{}_i{}^j[a_l],
\end{equation}
where each $C^{(n)}[c,a_l]$ is a local functional 
of $c$ and $a_l$; 
the gauge potential $A^{(n)}_i[a_l]$ and the vielbein 
$E^{(n)}{}_i{}^j[a_l]$ are local functionals of 
the commutative gauge field $a_l$. 
Furthermore we require for the zeroth order contributions
\begin{equation}
  C^{(0)}=c,\quad A^{(0)}{}_i=a_i,\quad E^{(0)}{}_i{}^j
  =\delta_i^j,
\end{equation}
which implies, that we obtain the usual commutative 
gauge theory in the limit $h\to 0$.

Solving Eqns. (\ref{sw-ghost}) and (\ref{sw-map-eqns}) 
one finds, that the ghost field $C$, the gauge field 
$A_i$ and the vielbein $E_i{}^j$ have to take values in 
the enveloping algebra of the Lie-algebra of the 
gauge group \cite{JSSW}.

It turns out, that at each level the Eqns. (\ref{sw-ghost}) 
and (\ref{sw-map-eqns}) can be written in the form 
(e.g. for the ghost field)\footnote{Notice, that this 
$\De$ is different from the $\De$ (coproduct) 
used in section \ref{funcReal}. There should be no 
confusion, since the context is different.}
\begin{equation}
  \label{deltaCm}
  \De C^{(m)}=Z^{(m)},\qquad\De\cdot\equiv s\cdot-i[c,\cdot]
\end{equation}
with $[\cdot,\cdot]$ being the graded commutator. 
$Z^{(m)}$ depends on the lower order terms  $C^{(n)}, n<m$ 
(as do $A^{(n)}{}_i$ and $E^{(n)}{}_i{}^j$ respectively) only. 
The operator $\Delta$ is a graded derivation, its 
square is zero:
\begin{equation}
  \De^2=0.
\end{equation}
Therefore for Eqn. (\ref{deltaCm}) to possess a solution 
the following consistency relation must hold:
\begin{equation}
  \De Z^{(m)}=0.
\end{equation}
This is indeed the case, as was shown for the ghost field 
in \cite{quadri}. By a similar calculation this can also 
be shown for $A_i$ and $E_i{}^j$, which should fulfill 
(\ref{sw-map-eqns}). However, one now has to take into 
account, that the operators $\ck\pat_i$ and $B_i{}^j$ 
are power series in the deformation parameter $h$.

The existence of the Seiberg-Witten map depends 
therefore on the triviality of the respective 
cohomology classes of the coboundary operator 
$\Delta$. Since in \cite{zumino} a homotopy 
operator ($K$, $K\Delta+\Delta K=1$) was 
constructed, this is indeed the case. 

The cohomological approach makes it clear, that 
there is at each order some freedom in the 
solution of the Eqns. (\ref{sw-ghost}), 
(\ref{sw-map-eqns}); namely, if $C^{(m)}$ is a 
solution of (\ref{deltaCm}), then 
$\tilde C^{(m)}=C^{(m)}+\Delta H$, with $H$ an 
arbitrary expression of the gauge field $a_i$, 
is also a solution.

For the example of the 2-dimensional quantum plane we obtain 
the following solutions to first nontrivial order \cite{diss}. 
The noncommutative gauge parameter has to fulfill
\begin{equation}
  \Delta C^{(1)}=-ix^1x^2\pat_2c\pat_1c;
\end{equation}
which has as solution
\begin{equation}
  \label{C-solution}
  C=c+h\frac{i}{2}x^1x^2\left[(\pat_2c)a_1-a_2(\pat_1c)\right]
  +{\cal O}(h^2).
\end{equation}

In the case of the vielbein and the gauge potential one 
has to expand the derivates $\ck\pat_i$ and the operators 
$B_i{}^j$ using the expressions (\ref{Man-*Wirk}) as a 
power series in $h$. E.g. $E^{(1)}{}_1{}^1$ must fulfill
\begin{eqnarray}
  \Delta E^{(1)}{}_1{}^1=-i(2x^1\pat_1c+x^2\pat_2c).
\end{eqnarray}
Which is easily solved:
\begin{eqnarray}
  \label{SW-E}
  E_1{}^1&=&1-ih(2x^1a_1+x^2a_2)+{\cal O}(h^2)\nn\\
  E_1{}^2&=&-ih2x^2a_1+{\cal O}(h^2)\nn\\
  E_2{}^1&=&0\nn\\
  E_2{}^2&=&1-ih(x^1a_1+2x^2a_2)+{\cal O}(h^2)
\end{eqnarray}
Using the ansatz
\begin{equation}
  \label{SW-A}
  A_i=a_i+hA_i^{(1)}+{\cal O}(h^2)
\end{equation}
and the first order solution (\ref{C-solution}) of the ghost 
field $C$, one obtains as possible solution to the second 
of the Eqns. (\ref{sw-map-eqns}):
\begin{eqnarray}
  A_1^{(1)}&=&(2x^2\pat_2+x^1\pat_1)a_1+i2x^2a_1a_2
  -\frac{i}{2}x^2a_2a_1+ix^1a_1a_1\nn\\
  &&+\frac{i}{2}x^1x^2\big[-f_{12}a_1
  +\pat_2a_1a_1-a_2\pat_1a_1\big]\nn\\
  A_2^{(1)}&=&(x^1\pat_1+x^2\pat_2)a_2
  +\frac{i}{2}x^1a_2a_1+ix^2a_2a_2\nn\\
  &&+\frac{i}{2}x^1x^2\big[-a_2f_{12}
  -a_2\pat_1a_2+\pat_2a_2a_1\big].
\end{eqnarray}
Where we used the commutative field strength 
$f_{12}=\pat_1a_2-\pat_2a_1-i[a_1,a_2]$. 

\section{Derivatives vs. Coordinates}

For a general noncommutative space it is possible 
to formulate gauge theory by using covariant 
coordinates \cite{MSSW,extradim}. Therefore 
one might ask, if there is a connection between 
the formulations using derivatives and coordinates 
respectively. 

For example, in the case of the canonical 
noncommutative structure
\begin{equation}
  [\x^i,\x^j]=i\theta^{ij},\qquad \theta^{ij}\in\mathbb{C}
\end{equation}
it is easily seen, that, for $\theta^{ij}$ invertible, 
the identification 
\begin{equation}
  \hat\pat_i=-i\theta_{ij}\x^j,\qquad\mbox{with }\;
  \theta_{ij}=(\theta^{-1})_{ij}
\end{equation}
is algebraical consistent, if one allows for 
noncommuting derivatives 
$[\hat\pat_i,\hat\pat_j]=i\theta_{ij}$. 
The respective gauge fields are then simply related by 
\begin{equation}
  A^i_{\mbox{\tiny coord}}=i\theta^{ij}
  A_{\mbox{\tiny deriv}, j}.
\end{equation}

At least for some $q$-deformed quantum spaces such 
relations between derivatives and coordinates also 
exist; but they are more complicated.
These relations include in general also the scaling 
operator $U$ and the generators $\lf^{\pm}_{ij}$ of 
the deformed enveloping algebra $U_q(g)$, and 
furthermore the algebra must be extended by the 
inverses of certain of its elements.

For $q$-deformed Euclidean spaces (i.e. spaces covariant 
under $SO_q$) it is possible to express the 
differential $df$ of a function as a commutator \cite{HS_int}:
\begin{equation}
  (1-q)df=[\omega,f]\equiv\frac{q^2}{1+q^2}
  \left[\hat r^{-2}d(\hat r^2),f\right].
\end{equation}
Where $\hat r^2=g_{ij}\x^i\x^j$ is the $q$-deformed 
length, which is central in the algebra of the 
coordinates and (co-)invariant under the action 
of the quantum group. Its differential can be 
written as $d(\hat r^2)=D_{kl}d\x^k\x^l$, with 
$D_{kl}=(1+q^{-1})g_{kl}$. Now, assuming that there 
are some operators $C_i^l$ such that 
$f(\x)\x^l\equiv \x^iC_i^lf(\x)$, we obtain
\begin{equation}
  [\omega,f]=\frac{1}{1+q}d\x^k\left[D_{kl}
    \frac{\x^l}{\hat r^2}-\frac{1}{\hat r^2}D_{ml}
    \x^iC_i^lB_k{}^m\right]f,
\end{equation}
which implies
\begin{equation}
  \hat\pat_k=\frac{1}{1-q^2}\frac{1}{\hat r^2}
  \left(D_{kl}\x^l-D_{ml}\x^iC_i^lB_k{}^m\right).
\end{equation}

Even for our example, the Manin plane, which is not Euclidean, 
one can read off the following identifications from Eqns. 
(\ref{man-abl-wirk}):
\begin{eqnarray}
  \ck\pat_1&=&+\la^{-1}\ck V^2\ck\lf^-_{11}\ck\lf^+_{12}
  \frac{1}{\ck x^2}\nn\\
  \ck\pat_2&=&-\la^{-1}\ck V^2\ck\lf^+_{11}\ck\lf^-_{21}
  \frac{1}{\ck x^1}.
\end{eqnarray}
Here we used $\ck V\equiv \ck U^{\frac{1}{3}}$. 
Algebraical consistency 
($\ck\pat_1\ck\pat_2=q^{-1}\ck\pat_2\ck\pat_1$) 
would however require an additional relation between the 
$L$-functionals and the coordinates.
This requirement is fulfilled, if one expresses the operators 
$\ck\lf^+_{12}$ and $\ck\lf^-_{21}$ as follows:
\begin{eqnarray}
  \ck\lf^+_{12}&=&-q^{-2}(\ck\lf^+_{22}-q^2\ck\lf^+_{11}
  \ck V^2)\ck x^2\frac{1}{\ck x^1}\nn\\
  \ck\lf^-_{21}&=&-q^2(\ck\lf^+_{22}-q^{-2}\ck\lf^+_{11}
  \ck V^{-2})\ck x^1\frac{1}{\ck x^2}.
\end{eqnarray}

\section{Remarks and Conclusions}

Of course, the whole gauge theory can be formulated 
by using directly differential forms \cite{DiMad}. 
As usual the covariant derivation and the gauge potential 
are then given by 
\begin{equation}
  D\psi\equiv d\psi+A\psi,\qquad A\equiv d\x^iA_i.
\end{equation}
The gauge potential is transforming according to
\begin{equation}
  \delta A=d\La+i[\La,A].
\end{equation}
Which agrees with (\ref{gauge-trafo}), if one takes 
into account that $\La d\x^i=d\x^j(B_j{}^i \La)$. 
For the field strength one has
\begin{equation}
  iF=dA-iAA,\qquad \delta F=i[\La,F].
\end{equation}
The torsion is related to the covariant derivation 
\begin{equation}
  De^j\equiv de^j-i[A,e^j]
\end{equation}
of the vielbein field $e^j=d\x^i e_i{}^j$.

However, if one works with $*$-products and has given a trace 
of the algebra (which we denote as $\int$ and where 
trace means in particular $\int f*g=\int g*f$), 
it seems to be more natural to use the field 
strength as defined in (\ref{tensors}), since it transforms as 
$\delta F_{ij}=i[\La\ds F_{ij}]$ and therefore for example 
\begin{equation}
  \int \mbox{Tr}\; F_{ij}*F^{ij}
\end{equation}
is invariant under gauge transformations ($\mbox{Tr}$ is 
a trace over the generators of the gauge group). 
In this way gauge invariant theories can be constructed 
\cite{JMSSW}.

In any case, this example shows, that if one wants to use some 
sort of ``vielbein'' in a noncommutative geometry, this vielbein 
has to transform nontrivially under gauge transformations. 

Above, in section \ref{sect_swmap}, we constructed the 
Seiberg-Witten map for the vielbein requiring that it reduces 
to $1$ in the $h\to 1$ limit: $E^{(0)}{}_i{}^j=\delta_i{}^j$. 
One could assume a $x$-dependent zeroth order term: 
$E^{(0)}{}_i{}^j=\ve_i^j(x)$. In this case a Seiberg-Witten map 
also exists. E.g. a first order solution is: 
\begin{eqnarray}
  E^{(1)}{}_1{}^1&=&-i\ve_1^1(2x^1a_1+x^2a_2)
  -ix^1x^2(a_2\pat_1\ve_1^1-a_1\pat_2\ve_1^1)\\
  E^{(1)}{}_1{}^2&=&-i\ve_1^2(x^1a_1+2x^2a_2)
  -ix^1x^2(a_2\pat_1\ve_1^2-a_1\pat_2\ve_1^2)
  -i2x^2\ve_1^1a_1\nn\\
  E^{(1)}{}_2{}^1&=&-i\ve_2^1(2x^1a_1+x^2a_2)
  -ix^1x^2(a_2\pat_1\ve_2^1-a_1\pat_2\ve_2^1)\nn\\
  E^{(1)}{}_2{}^2&=&-i\ve_2^2(x^1a_1+2x^2a_2)
  -ix^1x^2(a_2\pat_1\ve_2^2-a_1\pat_2\ve_2^2)
  -i2x^2\ve_2^1a_1\nn
\end{eqnarray}

For the action of the matter fields one needs also fields 
$\overline\psi$ transforming according to 
$\de\overline\psi=-i\overline\psi\La$:
\begin{equation}
  \de\overline\psi=-i\overline\psi\La,\qquad 
  {\cD}_i\overline\psi=E_i{}^j\left(\pat_j\overline\psi
    +i(B_j{}^k\overline\psi)A_k\right).
\end{equation}
A Seiberg-Witten map can be constructed along the above lines. 

It would be interesting, to know if one can express the gauge 
field in terms of the vielbein (or the vielbein in terms of 
the gauge field). Inverting the 
Seiberg-Witten map for $A_i$ (\ref{SW-A}) and inserting 
$a_i[A_j]$ in the expression for the vielbein (\ref{SW-E}) this 
seems to be possible. 
Also the question arises, whether there are gauge and vielbein 
fields without torsion, apart from pure gauge fields. 
The Seiberg-Witten map does not give torsion free fields unless 
the commutative field strength $f_{12}$ vanishes:
\begin{equation}
  T_{12}{}^1=ihx^1f_{12}+{\cal O}(h^2),
  \qquad T_{12}{}^2=2ihx^2f_{12}+{\cal O}(h^2).
\end{equation}

Furthermore, our results demonstrate, at least for special cases, 
that for algebras corresponding to a nonconstant Poisson-tensor 
$\theta^{ij}$ it is possible to construct covariant derivatives. 
Also some sort of Seiberg-Witten map exists. 
Having in mind general coordinate transformations, which clearly 
lead to nonconstant Poisson-structures, it would be interesting to 
know, if there exists a formulation which works 
for the generic case and uses derivatives. 
Or, if one has to go back to covariant 
coordinates, which always exist, but don't possess a nice 
commutative limit.

\section*{Acknowledgements}

I would like to thank J. Wess for useful discussions, B. Zumino for 
comments on the manuscript and the Theoretical Physics Group of the 
Lawrence Berkeley National Laboratory for kind hospitality. 

This work was supported by the Max-Planck-Society, Germany.
This work was supported in part by the Director, Office of
Science, Office of High Energy and Nuclear Physics, 
Division of High 
Energy Physics of the U.S. Department of Energy under Contract 
DE-AC03-76SF00098 and in part by the
National Science Foundation under grant PHY-0098840.


\end{document}